\documentclass[aps,prl,superscriptaddress,twocolumn,balancelastpage,nofootinbib
]{revtex4-2}
\usepackage[colorlinks,bookmarks=false,citecolor=blue,linkcolor=blue,urlcolor=blue]{hyperref}
\usepackage[all]{hypcap}   

\usepackage{amsmath,amssymb}
\usepackage{graphicx}

\usepackage{verbatim}
\usepackage{color}

\usepackage{placeins}    
\usepackage{flafter}     
\usepackage{color}

\newcommand{\ui}{\text{i}}

\newcommand{\diff}{\ensuremath{\mathrm{d}}}

\providecommand*{\oneD}{\textsc{1d}}
\providecommand*{\twoD}{\textsc{2d}}

\providecommand*{\threeD}{\textsc{3d}}

\providecommand*{\fourD}{\textsc{4d}}

\newcommand{\eqnref}[1]{Eq.~(\ref{#1})}
\newcommand{\figref}[1]{Fig.~\ref{#1}}
\newcommand{\Figref}[1]{Figure~\ref{#1}}

\newcommand{\Phieff}{\ensuremath{\Phi_{\text{eff}}}}

\makeatletter
\let\Hy@backout\@gobble
\makeatother

\begin{document}
\title{Quantum transport through partial barriers in higher-dimensional systems}

\author{Jonas St\"ober}
\affiliation{TU Dresden,
 Institute of Theoretical Physics and Center for Dynamics,
 01062 Dresden, Germany}

\author{Arnd B\"acker}
\affiliation{TU Dresden,
 Institute of Theoretical Physics and Center for Dynamics,
 01062 Dresden, Germany}

\author{Roland Ketzmerick}
\affiliation{TU Dresden,
 Institute of Theoretical Physics and Center for Dynamics,
 01062 Dresden, Germany}

\date{\today}

\begin{abstract}
Partial transport barriers in the chaotic sea of Hamiltonian systems
influence classical transport, as they allow for a small
flux between chaotic phase-space regions only. We establish for higher-dimensional
systems that quantum transport through such a partial barrier follows a
universal transition from quantum suppression to mimicking classical
transport.
The scaling parameter involves the flux,
the size of a Planck cell, and the localization length due to dynamical
localization along a resonance channel.
This is numerically demonstrated for coupled kicked rotors with a partial
barrier that generalizes a cantorus to higher dimensions.
\end{abstract}

\maketitle


\begin{figure}
 \includegraphics{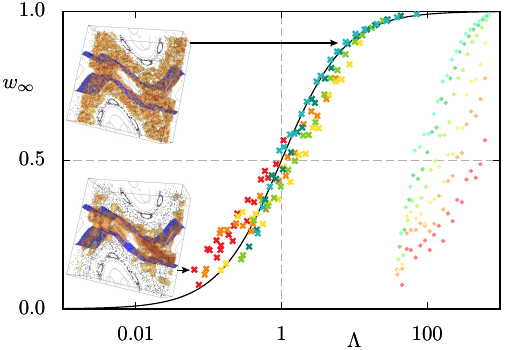}
 \caption{Universal transition from quantum suppression to classical transport through a partial barrier:
Dark-colored crosses show the transmitted weight $w_\infty$, \eqnref{eq:transmitted_weight},
vs.\ scaling parameter $\Lambda=\Phieff/h^2$, \eqnref{eq:scaling_parameter}, compared to \eqnref{eq:transition_curve} (solid line).
Light-colored circles correspond to the incorrect
scaling $\Lambda=\Phi/h^2$. The system is the kicked Hamiltonian, \eqnref{eq:hamiltonian}, with parameters given in the text.
The insets show
the Husimi distribution of a single time-evolved state, see \figref{fig:phase_space}.}
 \label{fig:transition}
\end{figure}
Many quantum phenomena can be well understood using quantum-classical correspondence. Of particular interest are situations where this is not the case and new quantum phenomena emerge. A famous example in the context of quantum chaotic transport is dynamical localization, where the quantum localization length is related to the classical diffusion coefficient
\cite{CasChiIzrFor1979, ChiIzrShe1981, FisGrePra1982, GrePraFis1984, CasChi1995:Collection, MooRobBhaSunRai1995, SeeMcCTanSuLuoZhaGup2022, CaoSajMasSimTanNolShiKonGalWel2022}.
Another example is quantum suppression of transport due to classical partial barriers, which is well studied in low-dimensional systems \cite{KayMeiPer1984b, BroWya1986, GeiRadRub1986, KayMei1988, BohTomUll1990a, SmiTomBoh1992, MaiHel2000, MicBaeKetStoTom2012}, but unexplored in higher dimensions.

A classical partial transport barrier separates two chaotic phase-space regions and allows for the exchange of a flux $\Phi$ between them \cite{KayMeiPer1984b, Mei1992, Mei2015}.
For example, in \twoD\ maps such partial barriers are \oneD\ lines and the most restrictive ones arise from invariant manifolds attached to hyperbolic points or from broken regular tori, so-called cantori. In a \fourD{} map, this has been recently generalized to a normally hyperbolic invariant manifold (NHIM) with the structure of a cantorus \cite{FirBaeKet2023}. This leads to a \threeD{} partial barrier characterized by a \fourD{} flux~$\Phi$. Note that in transition state theory \cite{TruGarKli1996, WaaWig2004, WaaSchWig2008, WaaWig2010, KayStr2014, EzrWig2018} for time-continuous Hamiltonian systems, partial barriers correspond
to dividing surfaces in phase space, e.g., separating reactants and
products of a chemical reaction.

Quantum transport for \twoD\ maps is suppressed if the size of a Planck cell $h$ is larger than the classical flux $\Phi$ \cite{KayMeiPer1984b, BroWya1986, GeiRadRub1986, KayMei1988, BohTomUll1990a, SmiTomBoh1992, MaiHel2000, MicBaeKetStoTom2012}. Furthermore, in \twoD\ maps there is a universal transition from quantum suppression to
mimicking classical transport depending on the scaling parameter $\Phi/h$ \cite{MicBaeKetStoTom2012}.
It is an open question whether such a quantum suppression also exists in higher-dimensional maps with $f\ge2$ degrees of freedom. If so, what is the scaling parameter? Can it be generalized from \twoD\ maps by comparing the classical flux $\Phi$ to the size $h^f$ of a Planck cell?
Finally, is the transition universally described by the same transition curve as for \twoD{} maps?

In this paper we establish a universal transition from quantum suppression to
mimicking classical transport through a partial barrier
in a \fourD\ map. We demonstrate that the scaling parameter is not $\Phi/h^2$, but that it has to be modified by the localization length of dynamical localization. This reveals for higher dimensions a remarkable combination of the two quantum phenomena of dynamical localization and quantum suppression of transport through a partial barrier.


\emph{Results.} --- The universal transition from quantum suppression to mimicking classical transport is demonstrated in \figref{fig:transition} for a \fourD\ kicked Hamiltonian, see \eqnref{eq:hamiltonian} below.
The transmitted weight $w_\infty$, defined in \eqnref{eq:transmitted_weight}, ranges from zero to one and measures the quantum transport through the partial barrier relative to the classical transport.
The transition is well described by the universal curve
\begin{equation}
    w_\infty = \frac{\Lambda}{1+\Lambda},
    \label{eq:transition_curve}
\end{equation} with the scaling parameter $\Lambda$.
In contrast to a \twoD\ map, where $\Lambda=\Phi/h$ \cite{MicBaeKetStoTom2012}, we find for a \fourD{} map,
\begin{equation}
    \Lambda = \frac{\Phieff}{h^2},
    \label{eq:scaling_parameter}
\end{equation}
where $h^2$ is the size of a Planck cell.
The effective \fourD\ flux is given by two factors,
\begin{equation}
    \Phieff=\lambda\;\Phi_{\threeD},
    \label{eq:effective_flux}
\end{equation}
which originate from the geometry and dynamics of the omnipresent resonance channels \cite{Arn1964,Chi1979,LicLie1992,OnkLanKetBae2016,GuzEftPae2020} in a \fourD\ phase space: (i) The chaotic region of a resonance channel extends in one direction and is enclosed by the partial barrier in all others.
Here $\Phi_{\threeD}$ is the local flux through the partial barrier at a point along the channel \cite{FirBaeKet2023}. (ii) The quantum localization length $\lambda$ is due to dynamical localization caused by classical diffusion along the one-dimensional channel \cite{CasChiIzrFor1979, ChiIzrShe1981, FisGrePra1982, GrePraFis1984, CasChi1995:Collection}.
Thus, the combination of classical and quantum properties allows for establishing a universal transition of quantum transport through a partial barrier in higher-dimensional systems.

\begin{figure}
 \centering
 \includegraphics{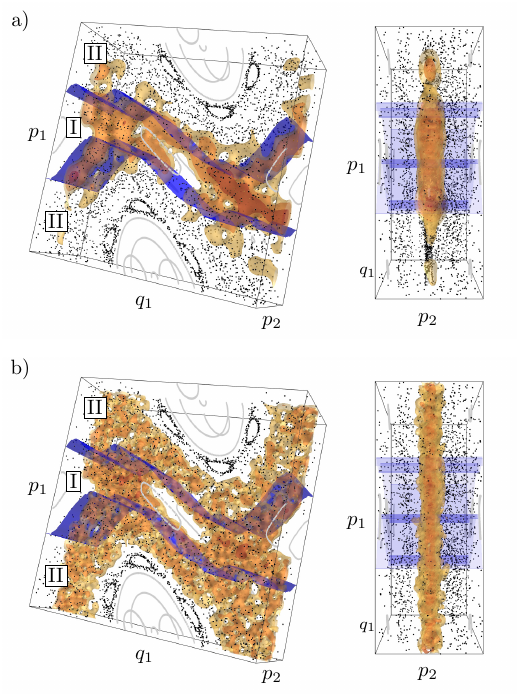}
 \caption{Phase space of the \fourD\ map, \eqnref{eq:map}, for $\xi=0.01$ represented by the \threeD\ phase-space slice $|q_2-0.5|<10^{-4}$ with $q_1, p_1\in[0.0, 1.0]$ and $p_2\in[-0.2, 0.2]$
with regular (gray) and chaotic (black) orbits. The partial
barriers (blue surfaces) associated to the cantorus-NHIM divide the
phase space into regions I and II.
The Husimi function of a single time-evolved quantum state (contours at 20\%, 2\% and 0.5\%) at $t=250000$ is shown in two perspective views for (a) $h=1/200$ and (b)
$h=1/800$.}
 \label{fig:phase_space}
\end{figure}

\emph{Classical system with a partial barrier.}---
To investigate the classical and quantum transport through a partial barrier we consider two coupled kicked rotors obtained from the kicked Hamiltonian
\begin{equation}
 H =T(p_1, p_2) + V(q_1, q_2)
 \sum_{n \, \in \, \mathbb{Z}} \delta(t - n)\label{eq:hamiltonian},
\end{equation}
with kinetic energy $T$ and kick potential $V$,
\begin{align}
  T(p_1, p_2)&=\frac{p_1^2}{2} + \frac{p_2^2}{2}\\
  V(q_1, q_2) &=\frac{K_1}{4 \pi^2}\cos{(2 \pi q_1)} + \frac{\xi}{4
  \pi^2}\cos{(2 \pi [q_1+q_2])},
\end{align}
where $K_1$ is the kicking strength and $\xi$ the coupling between the degrees
of freedom. Stroboscopically, after each kick, one obtains a \fourD\ symplectic map \cite{Fro1972},
\begin{subequations}
  \begin{align}
    q'_1 &= q_1 + p_1\\
    q'_2 &= q_2 + p_2\\
    p'_1 &= p_1 + \frac{K_1}{2\pi}\sin(2\pi q'_1) + \frac{\xi}{2\pi}\sin(2\pi
    [q'_1+q'_2])\\
    p'_2 &= p_2 + \frac{\xi}{2\pi}\sin(2\pi[q'_1+q'_2]).
  \end{align}
  \label{eq:map}%
\end{subequations}%
Periodic boundary conditions with period 1 are imposed.
In the following we fix the kicking strength $K_1=1.4$.

For the uncoupled case
$\xi=0$, the first degree of freedom ($q_1,p_1$) corresponds
to a single kicked rotor. The chaotic dynamics in $p_1$-direction is restricted by one-dimensional partial transport barriers extending in $q_1$-direction. The flux $\Phi$ is the \twoD\ phase-space area which is transported across the partial barrier in one application of the map. The partial barrier with the smallest flux is based on the golden cantorus, which is a broken invariant curve of golden mean frequency \cite{Per1982,KayMeiPer1984b,Mei1992,Mei2015}.
This cantorus is extended by the uncoupled second degree of freedom ($q_2$, $p_2$) to a so-called cantorus-NHIM \cite{FirBaeKet2023} --- a normally hyperbolic
invariant manifold \cite{Fen1972,Wig1994,Eld2013} with the structure of a cantorus. This gives rise to a \threeD\ partial barrier in the \fourD\ phase space with \fourD\ flux $\Phi$ \cite{FirBaeKet2023}.

For small couplings $\xi>0$, the cantorus-NHIM persists
and is slightly deformed as well as the partial barrier and the \fourD\ flux changes. The coupling leads to a slow diffusion in $p_2$-direction so that for transport through the partial barrier a local flux $\Phi_{\threeD}(p_2)$ becomes relevant \cite{FirBaeKet2023}.

In \figref{fig:phase_space} we visualize the phase space of the \fourD\ map,
\eqnref{eq:map}, by means of a \threeD\
phase-space slice \cite{RicLanBaeKet2014}.
The partial barrier is a \threeD\ surface in the \fourD\ phase space so that it is a \twoD\ surface (blue) in the \threeD\ phase-space slice.
Due to symmetry, there exist two partial barriers that divide
the phase space into two distinct regions I and II, see
\figref{fig:phase_space}.
Geometrically, each region forms a so-called resonance channel~\cite{Arn1964,Chi1979,LicLie1992,OnkLanKetBae2016,GuzEftPae2020}
which extend in $p_2$-direction and are composed of regular and chaotic motion.

\emph{Quantum dynamics and transmitted weight.}---
Quantum mechanically the dynamics of
the kicked Hamiltonian \eqref{eq:hamiltonian} is described by a Floquet operator
\begin{equation}
U=\exp\left(-\frac{\ui}{\hbar} V\right) \exp\left(-\frac{\ui}{\hbar} T\right)
\end{equation} acting on an
$N^2$-dimensional Hilbert space where $h=1/N$ is the effective Planck constant, see ~Refs.~\cite{RivSarAlm2000, Lak2001,RicLanBaeKet2014}.
The time evolution of a state is given by
$|\psi(t+1)\rangle=U|\psi(t)\rangle$ and can be investigated experimentally \cite{GadReeKriSch2013}.

We want to study the time evolution
of states initially placed in region I between the symmetry-related partial barriers
and determine the transport through these partial barriers to region II, see \figref{fig:phase_space}. To quantify transport we consider a phase-space volume $B$ in region II and a projection operator
$P_B$ onto $B$. The weight of a state in $B$ at some time $t$ is
given by $\mu_B[\psi(t)]=\langle\psi(t)|P_B|\psi(t)\rangle$. This can be compared to the classical measure $\mu_B^{\text{ch}}=V^{\text{ch}}_B/V^{\text{ch}}$, where $V^{\text{ch}}$ is the total chaotic phase-space volume and $V_B^{\text{ch}}$ the chaotic phase-space volume in $B$. This leads to the (relative asymptotic) \emph{transmitted weight} \cite{MicBaeKetStoTom2012}
\begin{equation}
  w_\infty[\psi] := \frac{\langle\,\mu_B[\psi(t)]\,\rangle_t}{\mu_B^{\text{ch}}}
  \label{eq:transmitted_weight}
\end{equation}
where $\langle\cdot \rangle_t$ is a time-average at large times.
It reaches $w_\infty=1$ if a time-evolved state spreads uniformly within
the chaotic region.
In contrast, if the time-evolved state does not enter $B$
one has $w_\infty=0$.

The transmitted weights $w_\infty$ are shown in \figref{fig:transition}
as light-colored circles using the naive scaling parameter $\Lambda=\Phi/h^2$.
No common transition curve is found.
Below we will discuss how universal scaling can be achieved.

For \figref{fig:transition} we choose initial states in the chaotic part of
region I that are composed of a coherent state $|\alpha(q_1, p_1)\rangle$ centered at $(q_1, p_1)$ in the first degree
of freedom and a momentum eigenstate at $p_2=0$ in the second degree of freedom, i.e.\
$|\psi(t=0)\rangle=|\alpha(q_1, p_1)\rangle\otimes|p_2=0\rangle$.
The phase-space volume $B$ for the measure $w_\infty$ is chosen to be a \fourD\ box with $p_1\in[-0.25, 0.25]$ and $q_1, q_2, p_2$ over the whole period.
It encloses a major part of region II and has the numerical advantage that the projector $P_B$ is a sum over the corresponding momentum eigenstates.
The time average in \eqnref{eq:transmitted_weight} is carried out starting at $t=250000$ over 1000 time steps.
The transmitted weight is further averaged over 6 different initial states with varying coordinates $(q_1, p_1)$ in the chaotic part of region I as well as 10 Bloch phases.
This is displayed in \figref{fig:transition} for 20 values of $h$ (ranging from $1/200$ to $1/800$) and coupling strengths~$\xi$
($0.001$, $0.005$, $0.01$, $0.02$, $0.03$, and $0.04$) using colors (red, orange, yellow, light green, dark green, and cyan).

The flux $\Phi$ varies just slightly from 0.00097 to 0.00109 when $\xi$ changes from 0.001 to 0.04. In contrast, the transmitted weight $w_\infty$ varies substantially, even for fixed $h$, indicating that there has to be some other mechanism that influences quantum transport.

\begin{figure}
 \centering
 \includegraphics{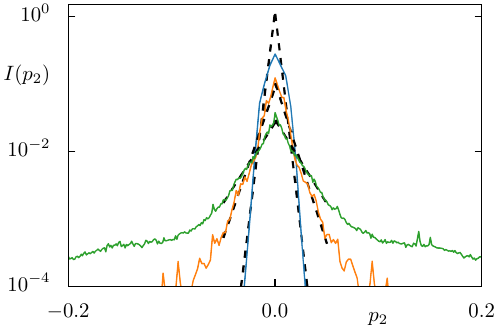}
 \caption{
 Intensity $I(p_2)$, \eqnref{eq:intensity_p2},
 of the
 time-evolved state
 started at $(q_1=0.2, p_1=-0.4)$ and $p_2=0$
 at $t=250000$ for
 $h=1/200$, $h=1/480$, and $h=1/800$
 (blue, orange, green, from narrow to wide).
   Exponential fits (dashed) determine the localization length $\lambda$.
   Note that the tails show a non-exponential decay at small values, which does not influence quantum transport through the partial barrier.
}
 \label{fig:diffusion}
\end{figure}

\emph{Dynamical localization and universal scaling.}---
In order to understand why the naive scaling parameter
$\Lambda=\Phi/h^2$ does not lead to a
common transition curve,
we consider the $p_2$-dependence of the intensity of the
time-evolved state,
\begin{equation}
  \label{eq:intensity_p2}
  I(p_2)=\int\;\diff p_1 |\psi(p_1, p_2)|^2
  \, ,
\end{equation}
shown in \figref{fig:diffusion}.
We find that the time-evolved state is exponentially localized in $p_2$.
This is due to dynamical localization~\cite{CasChiIzrFor1979, ChiIzrShe1981, FisGrePra1982, GrePraFis1984, CasChi1995:Collection}
caused by diffusion in $p_2$-direction.
Dynamical localization is generically expected in resonance channels of
higher-dimensional systems and has been demonstrated for \fourD\ maps~\cite{DemIzrMal2002a, DemIzrMal2002b}.
We determine the localization length $\lambda$
by fitting an exponential decay to the central part of $I(p_2)$ at sufficiently large times
(and additionally averaging over a time interval, initial states, and Bloch phases).
The localization length $\lambda$ increases with decreasing Planck constant $h$, as expected for dynamical localization \cite{CasChiIzrFor1979, ChiIzrShe1981, FisGrePra1982, GrePraFis1984, CasChi1995:Collection}.

This localization in $p_2$-direction
has an important consequence for describing
the transition from quantum suppression to mimicking classical transport.
The \fourD\ flux~$\Phi$ corresponds to a phase-space volume
extending along the entire $p_2$-coordinate.
For a localized time-evolved state,
however, we propose that
just an effective flux
$\Phieff = \lambda \, \Phi_{\threeD}$, Eq.~\eqref{eq:effective_flux},
is relevant.
Here $\Phi_{\threeD}$ is the local flux~\cite{FirBaeKet2023} through the partial barrier at the point $p_2$, where the time-evolved state
localizes.
For simplicity, we assume that the variation of $\Phi_{\threeD}$
within the localization length $\lambda$ is
negligible, so that the effective flux $\Phieff$
can be written as a product of $\lambda$ and $\Phi_{\threeD}$.
Therefore we use the scaling parameter $\Lambda=\Phieff/h^2$, Eq.~\eqref{eq:scaling_parameter}.

\Figref{fig:transition} shows that using this adapted scaling parameter the transmitted weights $w_\infty$
follow a universal curve.
The transition is well described by $\Lambda/(1+\Lambda)$,
Eq.~\eqref{eq:transition_curve},
previously found for a \twoD\ map~\cite{MicBaeKetStoTom2012}, and thus establishes universality for different phase-space dimensions.

The effective flux $\Phi_{\text{eff}}$ increases under variation of $\xi$ from 0.001 to 0.04 mainly due the localization length $\lambda$, which for example for $h=1/800$ increases from
0.002 
to 0.114. 
In contrast, $\Phi_{\threeD}(p_2=0)$ varies just slightly from 0.00097 
to 0.00098. 
This highlights the relevance of dynamical localization along a resonance channel for quantum transport through a partial barrier.

Note that due to the presence of two partial barriers one
would have to use twice the flux in the scaling parameter. However, this is compensated due to the symmetry of the map~\cite{Sto2023}.
Furthermore, we find that the universal scaling also holds for initial states with different values of $p_2$, which lead to substantial changes in the localization lengths \cite{Sto2023}.


\emph{Summary and Outlook.}---
We have established that there is a universal transition
from quantum suppression to mimicking classical
transport through a partial barrier in a \fourD{} map. The transition is well described by \eqnref{eq:transition_curve} which is the same as for lower-dimensional systems.
The universality is achieved when using the scaling parameter, \eqnref{eq:scaling_parameter},
involving the Planck cell of size $h^2$ and the effective flux $\Phieff$.
For this the localization length $\lambda$ due to dynamical
localization along a resonance channel plays a key role, see \eqnref{eq:effective_flux}.
This is demonstrated in Fig.~\ref{fig:transition}
for the transmitted weight through
a partial barrier associated with a cantorus-NHIM.

More generally, if in part of a resonance channel there is no dynamical localization, e.g.\ due
to drift-induced delocalization \cite{SchBaeKet2023:p}, the localization length
in \eqnref{eq:effective_flux} has to be replaced by the length of this part of
the resonance channel.
We expect that the universal transition also applies to even higher-dimensional maps with $f\ge3$ degrees of freedom, when using the scaling parameter $\Lambda=\Phi_{\text{eff}}/h^f$. This can be extended from time-discrete maps also to time-continuous Hamiltonian systems.
The universal transition is also reflected in the properties of the chaotic eigenstates which are either confined on one side of the partial barrier or extend beyond it.
Potential applications are in the context
of chemical reactions and in many-body quantum chaos.

\acknowledgments

We are grateful for discussions with Jan Robert Schmidt.
Funded by the Deutsche Forschungsgemeinschaft (DFG, German Research Foundation) -- 290128388.


\end{document}